\documentclass[aps,prl,floatfix,twocolumn]{revtex4}
\usepackage{amsmath}
\usepackage{latexsym}
\usepackage{graphicx}
\usepackage{color}
\usepackage{float}
\newfloat{suppfig}{tbh}{losf}
\floatname{suppfig}{\sc{Supplementary Figure}}

\newcommand{\dee}{\mathrm{d}}

\newcommand{\ket}[1]{|#1\rangle}

\begin{document}

\title{Spectral compression of single photons}

\author{Jonathan Lavoie$^1$, John M. Donohue$^1$, Logan G. Wright$^1$, Alessandro Fedrizzi$^2$, and Kevin J. Resch$^1$}
\affiliation{$^1$Institute for Quantum Computing and Department of Physics \&
Astronomy, University of Waterloo, Waterloo, Canada, N2L 3G1\\ $^2$Centre for Engineered Quantum Systems and Centre for Quantum Computer and Communication Technology, School of Mathematics and Physics, University of Queensland, Brisbane 4072, Australia}

\maketitle

\textit{Abstract} - \textbf{Photons are critical to quantum technologies since they can be
used for virtually all quantum information tasks: in quantum
metrology~\cite{higgins2007entanglement}, as the information
carrier in photonic quantum
computation~\cite{kok2007loq,aspuru2012photonic}, as a mediator
in hybrid systems~\cite{wallquist2009hybrid}, and to establish
long distance networks~\cite{duan2001long}. The physical
characteristics of photons in these applications differ
drastically; spectral bandwidths span 12 orders of magnitude
from 50 THz~\cite{sergienko_broadband} for quantum-optical
coherence tomography~\cite{QOCT_Th} to 50~Hz for certain
quantum memories~\cite{tittel2010photon}. Combining these
technologies requires coherent interfaces that reversibly map
centre frequencies and bandwidths of photons to avoid excessive
loss. Here we demonstrate bandwidth compression of single
photons by a factor 40 and tunability over a range 70 times
that bandwidth via sum-frequency generation with chirped laser
pulses. This constitutes a time-to-frequency interface for
light capable of converting time-bin to colour
entanglement~\cite{Pol_to_Color_Ramelow2009} and enables
ultrafast timing measurements. It is a step toward arbitrary
waveform generation~\cite{kielpinski} for single and entangled
photons.}

Coherent photonic interfaces are paramount for future quantum
technologies. In quantum networks~\cite{duan2001long} for
example, photon pairs at 1550~nm---optimum for low-loss
transmission---distribute entanglement between network nodes
consisting of quantum memories. Parametric downconversion
(SPDC) sources are widespread for producing entangled photon
pairs~\cite{review_SinglePhoton}, and typically yield spectral
bandwidths of 300~GHz. The most efficient quantum
memories~\cite{hosseini2011high} however typically operate in
the near visible wavelength regime near 800~nm with narrower
bandwidths on the order of 10~MHz. This dilemma has partly been
addressed through centre-frequency
conversion~\cite{Upconversion_Huang1992} of single photons
using nonlinear optical processes in crystals~
\cite{Upconversion_Kwiat2003,Upconversion_Gisin2005,langrock2005highly,ramelow2012pec,rakher2011_modulationDot},
photonic crystal fibres~\cite{FWM_McGuinness2010} and Rubidium
vapor~\cite{FWM_Dudin2010}. This conversion process can be
highly efficient~\cite{langrock2005highly,pelc1} and can
conserve quantum coherence~
\cite{Upconversion_Huang1992,Upconversion_Gisin2005,FWM_Dudin2010,ramelow2012pec,rakher2011_modulationDot,FWM_McGuinness2010}.
Some quantum memory schemes~\cite{memory2} offer very limited
control over the spectrum of reemitted photons through varying
parameters of the control laser~\cite{shaped_EIT}.

Nonlinear optics has much more potential for manipulating and
controlling the spectrum of single photons. Sum-frequency
generation (SFG) is a nonlinear optical process in which a pair
of optical fields of frequencies $\nu_1$ and $\nu_2$ create a
third field with frequency $\nu_3=\nu_1 + \nu_2$~\cite{Boyd}.
When the driving fields are transform-limited laser pulses and
the acceptance bandwidth of the material is sufficiently large,
the bandwidth of the SFG is larger than that of the input
fields. Repeating this process with shaped, rather than
transform-limited pulses, can drastically alter the spectrum of
the SFG signal. Specifically, oppositely-chirped laser pulses
in SFG and equally-chirped laser pulses in difference-frequency
generation lead to narrow output spectra
~\cite{BC_oppositechirp_1, BC_oppositechirp_2, BC_DFM_1}. It
was recently proposed theoretically to employ SFG between a
chirped classical pulse and a single photon to enable
compression of the photon pulse in time~\cite{kielpinski};
interaction of a short laser pulse with the emission from a
quantum dot achieves~\cite{rakher2011_modulationDot} similar
results through temporal gating. In this letter, we exploit
pulse chirping of a classical laser and a single photon to
compress the bandwidth of the single photon, from 1740~GHz to
43~GHz, which is nearing the bandwidth regime of some
quantum memories~\cite{memory1,walmsley1}. When combined with
shaped pulses, nonlinear optics in the quantum regime promises
a new level of control over single photons~\cite{kielpinski,
Silberhorn2011}.

\begin{figure*}
  \begin{center}
   \includegraphics[width=2\columnwidth]{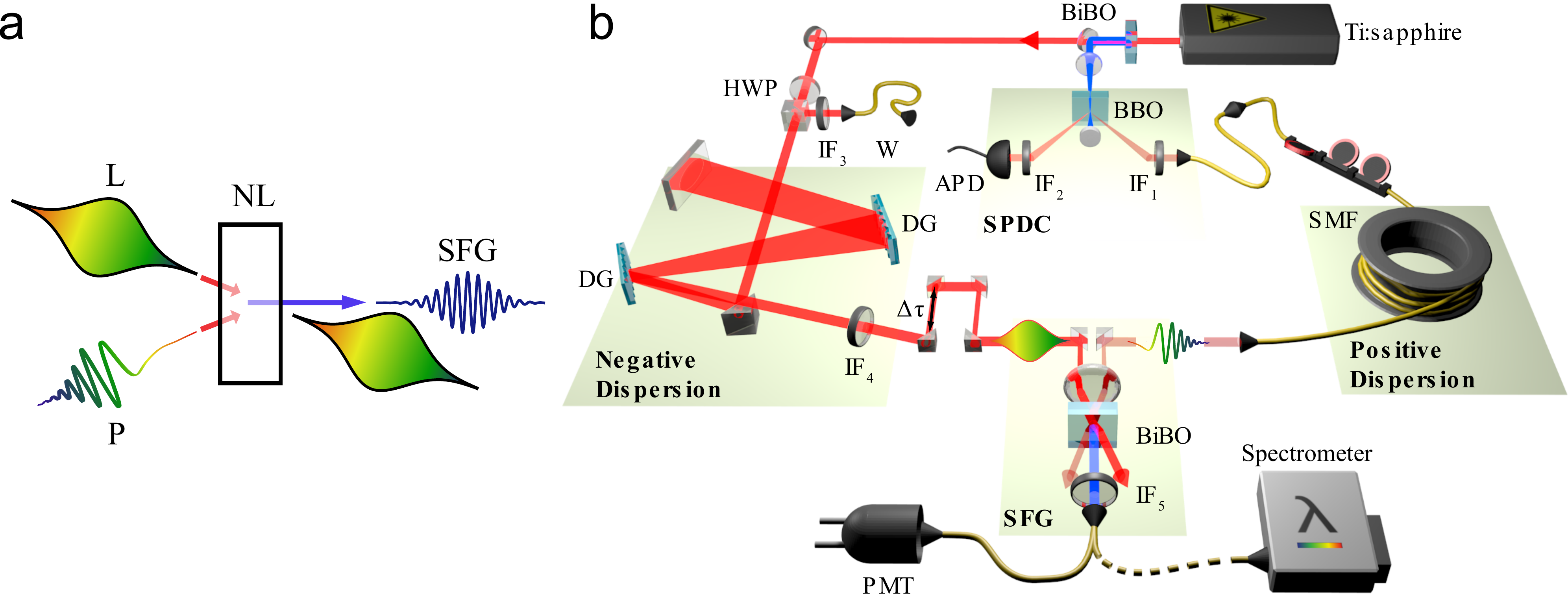}
  \end{center}
 \caption{ \textbf{Single-photon bandwidth compression scheme.} \textbf{a}, A broad-bandwidth single
 photon (P) with a linear frequency chirp is converted into a narrow-band
photon of a higher frequency via sum-frequency generation (SFG) with a strong laser pulse (L) of opposite chirp, in a
nonlinear crystal (NL). \textbf{b}, Experimental setup. The photon P is generated via spontaneous
parametric down-conversion (SPDC) in a $\beta$-Barium-Borate (BBO) crystal (Type-I, 1~mm) and sent through 34~m of optical
fiber (SMF) to introduce a linear chirp via
group velocity dispersion~\cite{Diels_Ultrashort}. The strong laser pulse is anti-chirped after a double-pass between two diffraction gratings (DG, 1200~lines\slash mm)~\cite{Diels_Ultrashort}. The spectrally narrowed photon is generated in a 1~mm thick Bismuth-Borate (BiBO) crystal and detected with a
photomultiplier tube (PMT) or sent to a spectrometer. For alignment
purposes, the single photons can be substituted by a weak coherent state (W), split off from the laser beam using a half-wave plate and a polarizing beam-splitter.\label{setup}}
\end{figure*}

Our scheme, depicted in Fig.~\ref{setup}(a), uses SFG between a
chirped single photon and an oppositely chirped intense laser
pulse.  The photon is chirped such that its frequency, centered
at $\nu_{0,P}$, increases in time, while the strong pulse
centred at $\nu_{0,L}$ is \emph{anti}-chirped such that its
frequency decreases linearly in time. When the photon and pulse
arrive at the crystal simultaneously, a red-shifted frequency
component ($\nu_{0,P}-\delta$) will meet a blue-shifted
component ($\nu_{0,L}+\delta$) with the same detuning $\delta$
and as a consequence, all frequency components will sum to a
narrow frequency centered on $\nu_{0,SFG}=\nu_{0,P}+\nu_{0,L}$.

A light pulse can be described by the frequency-dependent
electric field, $E(\nu)=U(\nu)e^{i\phi(\nu)}$, where $U(\nu)$
and $\phi(\nu)$ define the amplitude and phase, respectively. A
linear chirp results when a transform-limited pulse is subject
to a quadratic phase, $\phi(\nu) \sim A(\nu-\nu_0)^2$, with
$\nu_0$ the central frequency and $A$ is a constant. A chirp
increases the pulse duration and causes its instantaneous frequency
to vary linearly in time $\nu(t)=\frac{d\phi(t)}{dt}=\nu_0 \pm 2
\pi\frac{1}{2A} t$. When the chirped single photon (P) and
anti-chirped strong laser pulse (L) have a relative time delay
at the nonlinear crystal, $\Delta\tau$, the expected
upconverted frequency is
\begin{eqnarray}
\nu_{0,SFG}(\tau)=\nu_{0,P}+\nu_{0,L}+ 2\pi\frac{1}{2A}\Delta\tau\label{eq_tunability},
\end{eqnarray}
\noindent where we assume the pulses have equal and opposite
chirp, $\pm A$.  We consider the large chirp limit, where each
pulse is stretched many times its transform-limited duration,
i.e., $A^2\Delta\nu^4 \gg 1$, where $\Delta\nu$ is the full
width at half-maximum (FWHM) of the spectral intensity
distribution. We show in the Supplementary Material that the
expected intensity bandwidth (FWHM) of the upconverted single
photon is
\begin{eqnarray}
\Delta\nu_{SFG}^{TH}\approx\frac{\ln{4}}{A}\sqrt{\frac{1}{\Delta\nu_P^2}+\frac{1}{\Delta\nu_L^2}}.\label{expectedwidth}
\end{eqnarray}
\noindent Our technique thus compresses the spectral bandwidth
of the photon by a factor inversely proportional to the chirp
parameter $A$. By adjusting the delay between the pulses, the
central frequency $\nu_{0, SFG}(\tau)$ of the upconverted
photons can be tuned over some frequency range, approximately
$\sqrt{\Delta\nu_P^2+\Delta\nu_L^2}$ (see
Supplementary Information), limited by the spectra of the
initial single photon and laser pulse.

\begin{figure}
  \begin{center}
   \includegraphics[width=1\columnwidth]{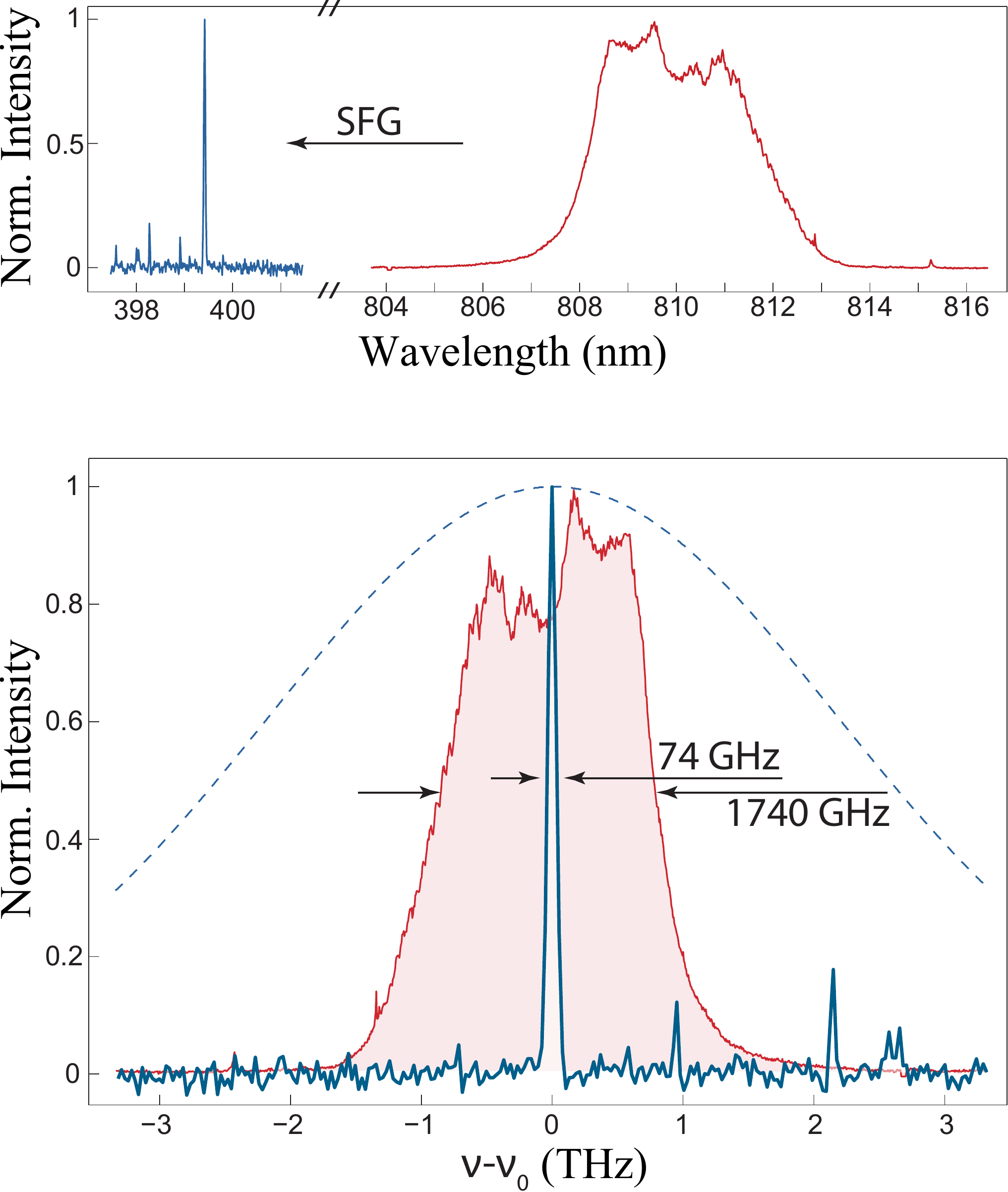}
  \end{center}
 \caption{\textbf{Single-photon spectra in wavelength (top) and relative frequency (bottom)}. The signal photons at the source (shown in red) have an initial bandwidth of 1740~GHz centred at 811~nm after transmission through an interference filter. Once the quadratic phase is applied and the photons are upconverted, the photon bandwidth reduces to $74\pm4$~GHz centred at 399.70~nm (blue curve). The spectra are shown as normalised spectral intensities and for the upconverted signal case, correspond to the average of six consecutive scans of $20$~minutes acquisition time. We  subtracted background counts determined by a supplementary scan with the signal photon path blocked. The blue dashed curve shows the theoretical spectrum of photons upconverted without our chirping technique but otherwise identical conditions.\label{spectrum}}
\end{figure}

The experimental setup is shown in Fig.~\ref{setup}(b); more
details are in the Methods. The second harmonic of a pulsed
laser is used to produce non-degenerate signal and idler
photons through down-conversion, while the remainder of the
fundamental serves as the strong classical pulse. We first measure the spectrum of the signal photons at the
source, after the interference filter IF$_1$, using a
fiber-coupled spectrometer and find a width
$\Delta\nu_P=1740\pm50$~GHz FWHM centered around
$811.11\pm0.01$~nm, see Fig.~\ref{spectrum}. The photons are
then sent through an optical fiber with positive dispersion and
superposed with the anti-chirped strong laser pulse
($\Delta\nu_L=4770\pm20$~GHz) centered at $787.62\pm0.02$~nm) at
the nonlinear crystal for sum-frequency generation.

The upconverted light is coupled into a single-mode fiber and
sent to the spectrometer. As shown in Fig.~\ref{spectrum}, we
observe significant spectral compression. The measured
bandwidth is $\Delta\nu_{M}=74\pm4$~GHz centered at 399.7~nm.
Taking the resolution of our spectrometer into account,
$\Delta\nu_R=60\pm4$~GHz (FWHM), the actual width of the
upconverted photon after deconvolution is
$\Delta\nu_{SFG}^{EXP}=\sqrt{\Delta\nu_{M}^{2}-\Delta\nu_R^2}=43\pm9$~GHz (see Supplementary Information for more details). This
agrees closely with theory, $\Delta\nu_{SFG}^{TH}=32.9\pm0.9$~GHz from equation~(\ref{expectedwidth}), using the expected
chirp parameter $A=(-25.8\pm0.3)\times10^6$~fs$^2$ given by the
geometry of our grating-based stretcher. We have therefore
achieved a compression ratio of 40:1 in the single photon
frequency bandwidth. Similar measurements were made after
replacing the single photons by a weak coherent state (W),
shown in Fig.~\ref{setup}(b). The corresponding measured
spectral width of the upconverted light was $67\pm4$~GHz, or
$30\pm12$~GHz after deconvolution, showing similar performance
with a classical chirped pulse with the same characteristics as
the signal photons.

Figure~\ref{spectrum} shows the spectrum predicted by theory
which would result from upconversion of the signal photon using
the same pump laser \emph{without} any chirp. The conversion in
this case actually \emph{broadens} the spectral bandwidth of
the original photon by a factor of 3, increasing the bandwidth
gap between flying broadband photons and narrowband quantum
memories, which further highlights the importance of our
scheme.

\begin{figure}
  \begin{center}
   \includegraphics[width=1\columnwidth]{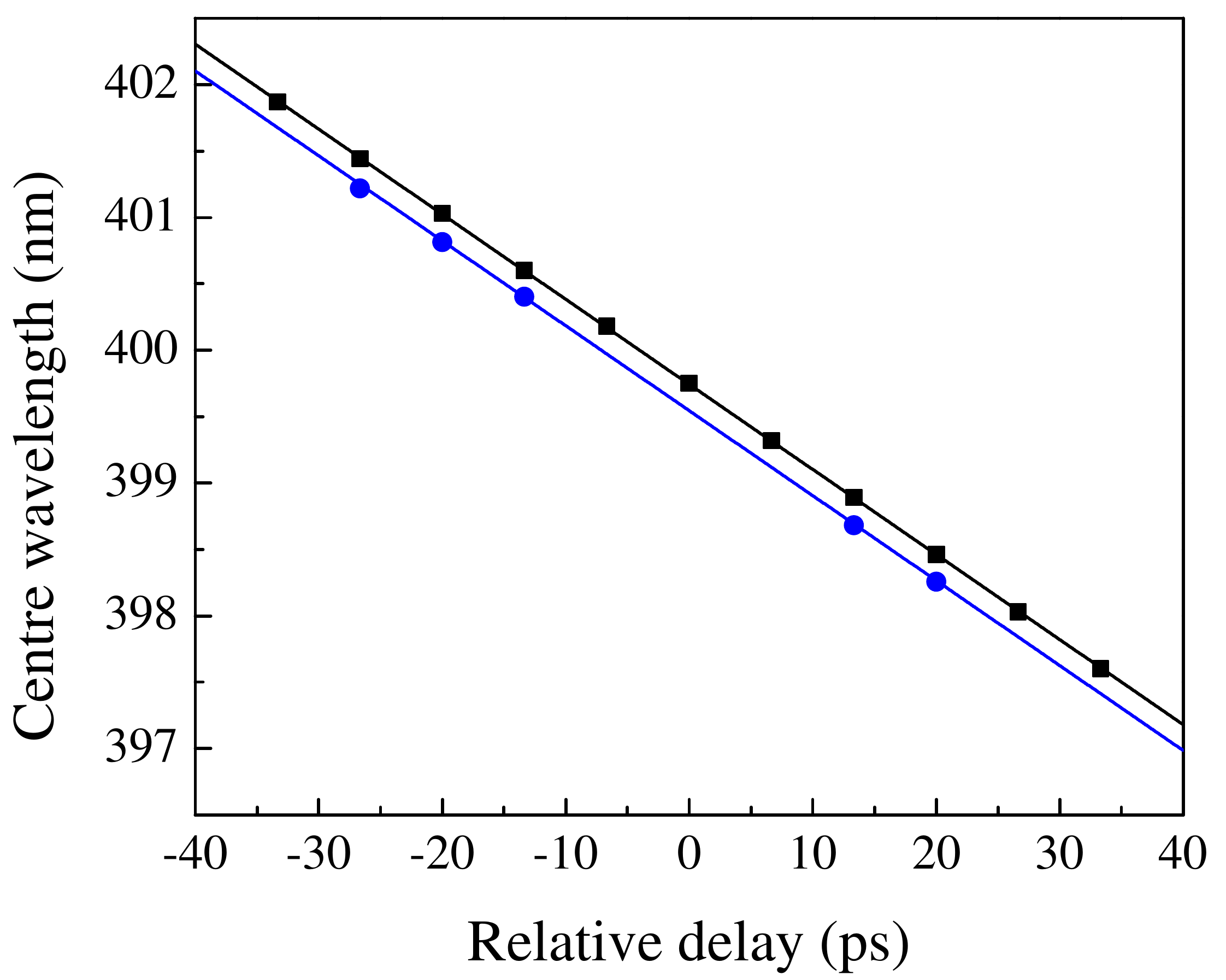}
  \end{center}
 \caption{\textbf{Wavelength tunability}. The center wavelength
of the upconverted light can be tuned by controling the relative
delay between the input pulses at the nonlinear crystal. The
blue circles represent the central wavelength of the upconverted single photon,
covering a range of 3~nm. The upconverted light from the weak coherent state behaves the same way and is plotted with
the black squares. The lines are linear fits yielding slopes of $-0.0640\pm0.0005$~nm/ps and $-0.0641\pm0.0001$~nm/ps for the single photons and the weak pulses, respectively. The vertical offset between the two curves ($\sim0.2$~nm) comes from a slight difference in the delay between the weak coherent state and the single photons at the crystal and a difference in their central wavelength. The error bars are smaller than the data points.\label{tunability}}
\end{figure}

The central wavelength of the narrowband upconverted photons
can be tuned by controlling the relative delay $\Delta\tau$
between the photon and the laser pulse, see
equation~(\ref{eq_tunability}). The SFG spectrum was measured
as a function of the delay of the single photon, with fitted
central wavelengths shown in Fig.~\ref{tunability}. The data
shows that the wavelength depends linearly on the delay as
expected. The linear fit gives a slope of $-0.0640\pm0.0005$~nm/ps in good agreement with $-0.0648\pm0.0008$ predicted from
equation~(\ref{eq_tunability}) and $A=(-25.8\pm0.3)\times10^6$~fs$^2$. If the weak coherent state is used instead of single
photons, we observe the same behaviour in the spectrum of the
upconverted light, and the data are also shown in
Fig.~\ref{tunability}. The lower SFG signal for single photons
required longer integration times than for the coherent states;
to reduce the effects of drift in experimental parameters all
of the data in Fig. 3 was taken within a day. For the single
photons, we focused on those delays demonstrating the widest
possible tuning range. From the slope, we can extract the chirp
parameter of $A=(26.2\pm0.2)\times10^6$~fs$^2$, in agreement
with expectations based on the parameters of the stretcher.

\begin{figure}
  \begin{center}
   \includegraphics[width=1\columnwidth]{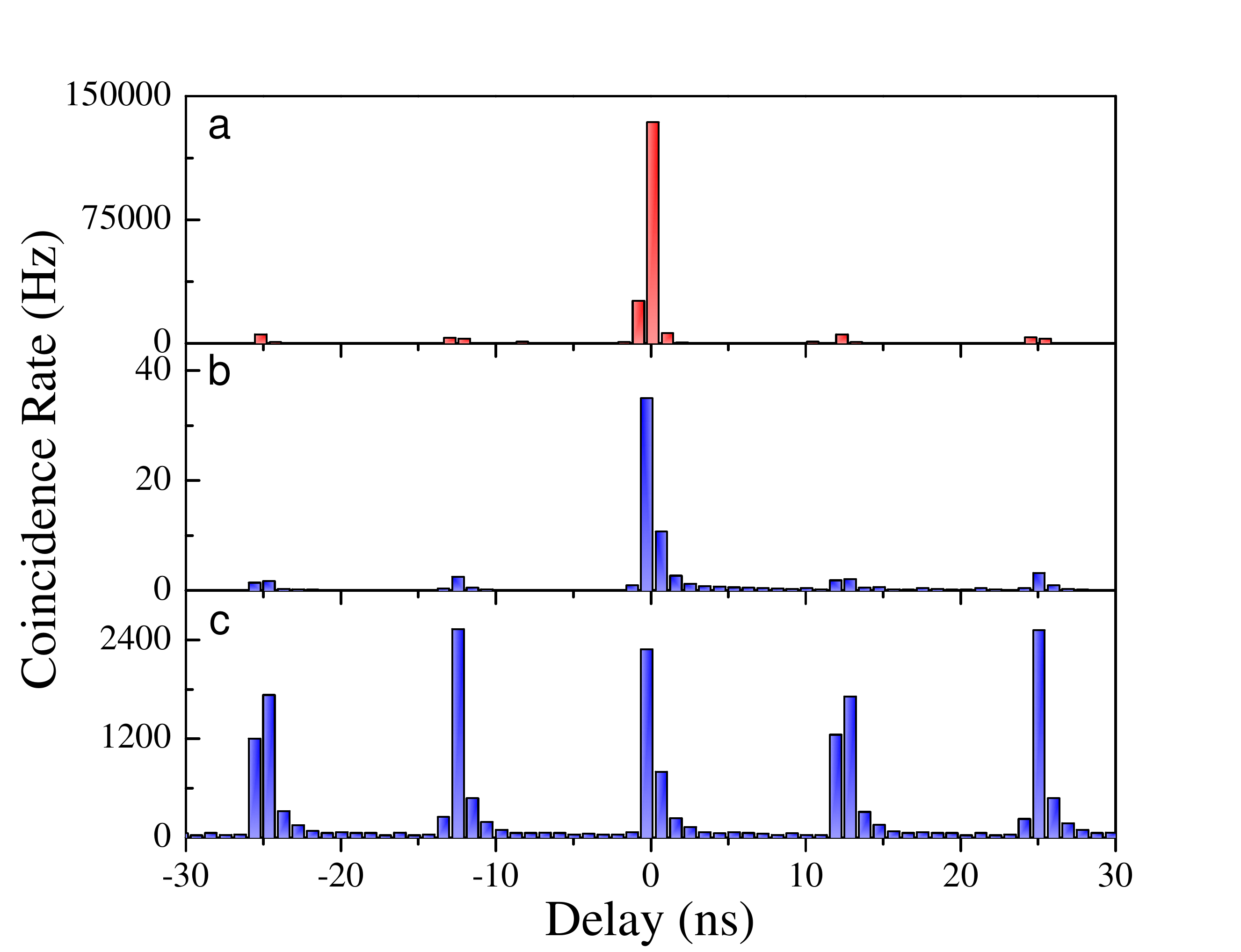}
  \end{center}
 \caption{\textbf{Temporal correlations with the idler photon}. \textbf{a}, The signal and idler from SPDC are produced in pairs, strongly correlated in time with a total measured coincidence rate of $160,000$~s$^{-1}$ around zero delay. \textbf{b}, The upconverted single photon maintains the strong timing correlation expected from individual photon pairs, and a coincidence rate of 50~s$^{-1}$ is detected. \textbf{c}, If the weak coherent state is upconverted instead, the histogram shows equal height peaks as expected for pulsed, but uncorrelated events. For each histogram, the optical path
 length difference with the idler is accounted for in post-processing and the abscissa is a variable electronic delay. An additional electronic delay box, with an observed asymmetric temporal jitter, was used on the idler side only, causing the asymmetry in b) and c). Error bars, 1 s.d. too small for this scale.\label{histograms}}
\end{figure}

Photons from SPDC are created in pairs and are strongly
correlated in time; we expect these correlations are preserved
through spectral compression. We first measured coincidence
counts between single photon detectors placed in the source as
a function of a time delay between the counts. Figure
\ref{histograms}(a) shows the coincidence counts versus delay,
without background substraction. The peak around zero delay
corresponds to a rate of 160,000~s$^{-1}$, within a 3~ns
coincidence window. The histogram also contains side peaks from
accidental coincidences with a separation of 12.5~ns matching
the laser repetition rate. After propagating through the
single-mode fibre, the signal photons are upconverted and
detected with a single-photon counting photomultiplier (PMT).
Figure \ref{histograms}(b) clearly shows that the coincidence
rate observed at zero-delay between the upconverted and idler
photons exceed the accidentals, preserving the strong timing
correlation associated with individual photon pairs. If we
instead upconvert the weak coherent state, which shares the
spectral and temporal properties of the signal photon, all
observed peaks in Fig.~\ref{histograms}(c) have the same
height, as expected for a pulsed, but temporally uncorrelated
source.

The efficiency of the upconversion process in our experiment is
$0.06\%$ with 300~mW of average power for the strong beam.
Dramatically higher efficiencies can be achieved with the use
of periodically poled crystals and higher pump powers
~\cite{langrock2005highly,pelc1}. The technique here can, in
principle, reach high efficiency; however higher laser power
will be required for increased compression. One doesn't require
$100\%$ conversion efficiency to achieve a net gain. Ignoring
the shift in centre frequency, one has an advantage once the
efficiency is greater than about $100\%/R$, where $R$ is the
ratio of the initial to final bandwidth, which in our case is
40, for an efficiency of just $2.5\%$. The amount of negative
dispersion achievable limits the maximum compression; with
achievable parameters~\cite{soyoBaek2009} and our spectra, one
could achieve compression down to $\sim$1~GHz.

Future work will explore the case when the single photon is
entangled with another. It is expected that polarization
entanglement could be preserved by employing
polarization-insensitive upconversion ~\cite{ramelow2012pec}.
Theoretical calculations show that energy-time entanglement
dramatically impacts this phenomenon (see Supplementary
Information). Our technique could serve as a coherent interface
between time-bin and frequency encodings of quantum information
due to the delay-dependent central frequency, allowing the
conversion of time-bin entanglement to colour
entanglement~\cite{Pol_to_Color_Ramelow2009}. It also enables
ultrafast timing measurements with slow
detectors~\cite{review_SinglePhoton} by converting different
pulse arrival times to different frequencies which could more
easily be distinguished. With our experimental parameters, one
could distinguish time bins with separation as short as 0.6~ps
over a 40~ps range. Future work will investigate nonlinear
interactions with more complex shaped pulses for manipulating
single photons; for example, coherent superpositions of chirped
pulses with different delays would allow ultrafast
single-photon time-bin measurements. The application of shaped
pulse nonlinear optics is a promising and unexplored regime at
the quantum level.

\section*{Methods}

Our laser (Spectra Physics Tsunami HP) has a 790~nm central wavelength,
10.5~nm spectral bandwidth (FWHM), 80~MHz repetition rate, and
2.5~W average power. The laser repetition rate is stabilized by a feedback system (Lok-to-clock)
which is important for the present experiment (see Supplementary Information). Second-harmonic generation in a $2$~mm
thick BiBO crystal yields a beam of $830$~mW, centred at
$394.2$~nm with 1.3~nm bandwidth, which generates the broadband
and non-collinear photon pairs. In the signal-photon path, the
interference filter IF$_1$ shown in Fig~\ref{setup}(b), with a
nominal bandwidth of 5~nm (FWHM) centered around 811~nm, is
inserted to keep its central frequency separated from that of
the laser beam at 790 nm, to separate the spectrum of the
second-harmonic background from our signal. Due to the energy
conservation of SPDC, an interference filter, IF$_2$, centered
around 770~nm and nominal bandwidth of 3~nm is inserted in the
path of the idler-photon.

We first optimize the setup with a weak coherent state (W),
split off from the strong laser pulse using a half-wave plate
and a polarizing beam-splitter. An interference filter (IF$_3$)
is inserted in its path and the transmitted spectrum has a
measured bandwidth of $3.85\pm0.03$~nm ($\Delta\nu_W=1760\pm10$~GHz) centered at $810.49\pm0.01$~nm, closely matched to IF$_1$.

The strong laser pulse was re-collimated before being
anti-chirped in a grating-based setup~\cite{Treacy_compressor},
with normal separation of 38~cm between the gratings, carefully
adjusted to minimize the spectral width of the upconverted
light. The bandpass filter IF$_4$ highly attenuates any power in
the tail of the laser spectrum above 800~nm to further suppress
second-harmonic background and after the filter, the pulse has a measured bandwidth
of $9.86\pm0.05$~nm ($\Delta\nu_L=4770\pm20$~GHz) centered at
$787.62\pm0.02$~nm. We
used an achromatic doublet with a 75~mm focal length in the
upconversion setup, and the conversion efficiency was optimized
by fine tuning the relative delay between the two inputs, the
spatial overlap at the crystal, and the phase matching by angle
tuning. A 14.5~nm bandpass filter IF$_5$ with central wavelength at 405~nm
is placed in the SFG beam to reduce
background.

The weak beam is finally replaced by the single photons, with
the same optical path length. We used a spectrometer (Acton
SP-2750A) with a 1200~lines/mm diffraction gratings blazed for
400~nm and entrance opening of 20$\mu$m with a resolution of 60~GHz. We use the spectrograph, free-running mode with a
back-illuminated CCD camera (PIXIS: 2048B).

\section*{Acknowledgements}

The authors thank Agata Branczyk, Deny Hamel, Ann Kallin, Roger Melko, Marco Piani, Robert Prevedel, Sven Ramelow, Krister Shalm, and John Watrous for helpful discussions. We are
grateful for financial support from the Natural Sciences and Engineering Research Council of Canada (NSERC), Ontario Centres of Excellence (OCE), the Canada Foundation for Innovation (CFI), QuantumWorks, the Ontario Graduate Scholarship Program (OGS) and the Ontario Ministry of Research and Innovation Early Researcher Award. A.F. is supported by an ARC Discovery Early Career Researcher Award DE130100240.

Correspondence and requests for materials
should be addressed to J.L. (e-mail: j3lavoie@uwaterloo.ca) or K.J.R.
(e-mail: kresch@uwaterloo.ca).

\clearpage

\renewcommand{\thesection}{S.\arabic{section}}
\renewcommand{\thesubsection}{\thesection.\arabic{subsection}}
\makeatletter 
\def\tagform@#1{\maketag@@@{(S\ignorespaces#1\unskip\@@italiccorr)}}
\makeatother
\makeatletter
\makeatletter \renewcommand{\fnum@figure}
{\figurename~S\thefigure}
\makeatother
\renewcommand{\figurename}{Figure}
\setcounter{equation}{0}

\section{SUPPLEMENTARY MATERIAL}

\section{THEORY}
We model the creation of upconverted single
photons through the interaction Hamiltonian,
$H$, of a three-wave $\chi^{(2)}$ non-linear process. We assume that frequency bandwidths are narrow and perfect phasematching ($\vec{k}_{P}+\vec{k}_{L}-\vec{k}_{SFG}=0$) is achieved, allowing us to simplify our Hamiltonian (ignoring constants) to
\begin{equation}
H\propto\iiint\mathrm{d}\nu_1\mathrm{d}\nu_2\mathrm{d}\nu_3\hat{a}^{(P)}_{\nu_1}\hat{a}^{(L)}_{\nu_2}\hat{a}^{\dag(SFG)}_{\nu_3}\delta(\nu_1+\nu_2-\nu_3)+h.c.
\end{equation}

Our initial three-mode state $\ket{\psi_i}$ is defined as a
single photon $\ket{1}_{P}=\int
d\nu_1f_P(\nu_1)e^{i\phi_P(\nu_1)}|1_{\nu_1}\rangle$,
a strong coherent state $|\alpha\rangle_{L}=\int
d\nu_2f_L(\nu_2)e^{i\phi_L(\nu_2)}|\alpha_{\nu_2}\rangle$
and an output signal, initially vacuum $\ket{0}_{SFG}$. For
each of the input pulses, the frequency distribution of the
field amplitude is assumed to be Gaussian,
$f_i(\nu_i)\propto\exp\left[-2\ln{2}\frac{(\nu_i-\nu_{0,i})^2}{\Delta\nu_i^2}\right]$,
with \emph{intensity} full-width at half-maximum (FWHM) $\Delta\nu_i$ and the phase term to
contain only linear and quadratic terms
$\phi_i(\nu_i)=2\pi(\nu_i-\nu_{0,i})\tau_i+A_i(\nu_i-\nu_{0,i})^2$,
where $\tau$ corresponds physically to a time delay and $A$ to
a linear chirp.

Assuming no frequency correlations between the single photons and idler, a lossless and nondispersive medium, and an
interaction time much longer than optical frequencies, the
first order term of the time evolution of the state (postselecting on the generation of a photon in the output mode) is found to
be proportional to
\begin{equation}\begin{array}{l}|\psi^{(2)}_f\rangle\propto\\ \iint
d\nu_3 d\nu_1
f_P(\nu_1)f_L(\nu_3-\nu_1)e^{i\left(\phi_P(\nu_1)+\phi_L(\nu_3-\nu_1)\right)}|1_{\nu_3}\rangle.\end{array}\end{equation}

From this, the electric field amplitude $E(\nu_3)$ as a function of frequency is thus proportional to
\begin{equation}E(\nu_3)\propto\int d\nu_1
f_P(\nu_1)f_L(\nu_3-\nu_1)e^{i\left(\phi_P(\nu_1)+\phi_L(\nu_3-\nu_1)\right)},
\end{equation}
which is the convolution of the frequency domain representation of the two input pulses. Classically, assuming narrow bandwidth, the field produced in sum-frequency generation is directly proportional to the product of the two input fields in the time domain, or the convolution of the fields in the frequency domain \cite{Boyd}. Thus, the same spectral properties are expected if both inputs are classical coherent states.

The field intensity $|E(\nu_3)|^2$ of the upconverted photons is found to have a spectral FWHM $\Delta\nu_{SFG}$ independent of time delay $\Delta\tau$,
\begin{equation}\begin{array}{l}\Delta\nu_{SFG}=\\\sqrt{\frac{(A_L+A_P)^2 \Delta\nu_L^4 \Delta\nu_P^4+4 \ln{4}(\Delta\nu_L^2+\Delta\nu_P^2)^2 }{(A_L^2\Delta\nu_L^2+A_P^2\Delta\nu_P^2)\Delta\nu_L^2\Delta\nu_P^2+4 \ln{4}(\Delta\nu_L^2+\Delta\nu_P^2) }}.\end{array}\end{equation}
In the case where no chirp is applied to the pulses, i.e. $A_P=A_L=0$, the bandwidth directly simplifies to $\sqrt{\Delta\nu_P^2+\Delta\nu_L^2}$. On the other hand, this bandwidth is minimized for equal and opposite chirps $A_P=-A_L=A$. If we also make the large chirp approximation that $A^2\Delta\nu^4\gg1$, the intensity width simplifies to \begin{equation}\Delta\nu_{SFG}\approx\frac{\ln{4}}{A}\sqrt{\left(\frac{1}{\Delta\nu_P^2}+\frac{1}{\Delta\nu_L^2}\right)},\end{equation} showing that the bandwidth is inversely proportional to the chirp rate. For our experimental parameters, $A^2\Delta\nu^4=6000$ and the approximation affects the expected frequency bandwidth by only 0.002\%.

The time delay $\Delta\tau$ appears as a shift in the central
frequency $\nu_{0,SFG}$ as well as an overall amplitude reduction, but has no fundamental effect on the bandwidth
of the signal. Making the same assumptions as before, the central
frequency $\nu_{0,SFG}$ is found to be a function of relative
delay $\Delta\tau\equiv\tau_2-\tau_1$ as
\begin{equation}\nu_{0,SFG}(\tau)=\nu_{0,P}+\nu_{0,L}+\frac{\pi}{A}\Delta\tau.\end{equation}
The efficiency of the upconversion process will depend
on the overlap of the two pulses. If we define the frequency
shift in equation (6) as $\delta\nu=\frac{\pi\Delta\tau}{A}$, the
intensity of the upconverted light contains a constant
prefactor reflecting the overlap,
$\exp\left[-\frac{4\ln{2}\delta\nu^2}{\Delta\nu_P^2+\Delta\nu_L^2}\right]$.
This prefactor limits the frequency tuning range to
$\sqrt{\Delta\nu_P^2+\Delta\nu_L^2}$ FWHM.

Converting to wavelength as in Fig.~3 in the
main text and expanding to a first-order Taylor series about
zero delay, the linear wavelength shift with delay $\Delta\tau$
is found to be
\begin{eqnarray}\lambda_{0,SFG}= & \frac{c
A\lambda_{0,P}\lambda_{0,P}}{c
A(\lambda_{0,P}+\lambda_{0,L})+\Delta\tau\pi\lambda_{0,L}\lambda_{0,P}}
\\ \approx &
\frac{\lambda_{0,L}\lambda_{0,P}}{\lambda_{0,P}+\lambda_{0,P}}-\frac{\pi\lambda_{0,P}^2\lambda_{0,P}^2}{
cA(\lambda_{0,L}+\lambda_{0,P})^2}\Delta\tau.\end{eqnarray}

\section{Effect of energy-time entanglement on bandwidth compression}
Spontaneous parametric down-conversion often emits photons with varying degrees of energy-time entanglement, depending on the phase-matching and pump characteristics. Thus, it is important to consider the effect of our compression scheme when the single photon is part of an energy-time entangled pair. As we show in the following calculations, the presence of energy-time entanglement significantly alters the achievable bandwidth compression with our scheme.

We assume that the photon of interest, the signal ($s$), is potentially entangled with a second photon, the idler ($i$), with a combined state of the form \begin{equation}\ket{\psi_{0}}\propto\iint \dee\nu_{s} \dee\nu_{i} f_P(\nu_{s},\nu_{i}) \ket{1_{\nu_{s}}}\ket{1_{\nu_{i}}}\ket{0}\end{equation} where $f_P(\nu_{s},\nu_{i})$ is the joint spectral amplitude and may be inseparable. Note that we implicitly assume a single spatial and polarizaton mode for each photon, and vacuum in a third SFG mode.

We consider a linear chirp $e^{iA(\nu_{s}-\nu_{0})^2}$ applied to the signal photon and use the Hamiltonian of equation (1) to calculate the sum-frequency generated by a nonlinear interaction with some anti-chirped strong laser pulse ($L$), whose spectral amplitude is $\alpha(\nu_2)=f_L(\nu_2)e^{i\phi_L(\nu_2)}$. Using first-order perturbation theory and only considering cases where a photon is created in the SFG mode, we have the state \begin{multline}\ket{\psi_f}\\ \propto \iiint\dee\nu_{s}\dee\nu_{i}\dee\nu_{\scriptscriptstyle SFG}F(\nu_{s},\nu_{i},\nu_{\scriptscriptstyle SFG})\ket{0}\ket{1_{\nu_{i}}}\ket{1_{\nu_{\scriptscriptstyle SFG}}}\end{multline} where $F(\nu_{s},\nu_{i},\nu_{\scriptscriptstyle SFG})=f_P(\nu_{s},\nu_{i})e^{iA(\nu_{s}-\nu_{0})^2}\alpha(\nu_{\scriptscriptstyle SFG}-\nu_{s})$. To find the spectral distribution of the upconverted photon, we trace out the idler photon, leaving \begin{multline}S(\nu_{\scriptscriptstyle SFG})\\ \propto\iiint\dee\nu_{s}\dee\nu_{s}'\dee\nu_{i}F^\ast(\nu_{s}',\nu_{i},\nu_{\scriptscriptstyle SFG})F(\nu_{s},\nu_{i},\nu_{\scriptscriptstyle SFG}).\end{multline}

To obtain some physical intuition, we employ the model two-photon spectral function \begin{multline}f_P(\nu_{s},\nu_{i}) \propto e^{-\frac{(\nu_{s}-\nu_{0})^2}{2\sigma^2}}e^{-\frac{(\nu_{i}-\nu_{0})^2}{2\sigma^2}}e^{-\frac{(\nu_{s}+\nu_{i}-2\nu_{0})^2}{2\sigma_c^2}}\end{multline} where $\sigma$ is related to the bandwidth of each photon. The constant $\sigma_c$, controls the degree of entanglement in the system. In the limit where $\sigma_c\rightarrow\infty$, the correlation term is constant and the spectral function is separable. In the opposite limit, where $\sigma_c\rightarrow 0$, the correlation term becomes a delta function $\delta(2\nu_0-\nu_s-\nu_i)$ and the two photons are perfectly energy-time entangled. For simplicity, we have assumed the photons are degenerate, where the central frequencies of the signal and idler photons are equal. Note that the intensity spectral bandwidth of one of the photons is $\Delta\nu_{P}=2\sqrt{\ln{2}}\sqrt{\frac{\sigma^2(\sigma_c^2+\sigma^2)}{\sigma_c^2+2\sigma^2}}$.

We further assume that no time delay is present between the pulses and the nonlinear crystal and that the chirps are equal and opposite, $A_s=-A_L=A$. Defining the intensity bandwidth (FWHM) of the strong laser pulse as $\Delta\nu_L=2\sqrt{\ln{2}}\sigma_L$, and taking the large chirp limit $A^2\sigma^4\gg1$, we find the intensity bandwidth (FWHM) of the SFG, {\small \begin{multline}\Delta\nu_{\scriptscriptstyle SFG} = 2\sqrt{\ln{2}}\times\\ \sqrt{\frac{[\sigma^2\sigma_L^2+\sigma_c^2(\sigma^2+\sigma_L^2)][\sigma^4+2\sigma^2\sigma_L^2+\sigma_c^2(\sigma^2+\sigma_L^2)]} {2\sigma^2\sigma_L^2\{\sigma^2+2A^2\sigma_c^2[\sigma^4+2\sigma^2\sigma_L^2+\sigma_c^2(\sigma^2+\sigma_L^2)]\}}}.\end{multline}}

We consider the extreme limits of $\sigma_c$ to draw some general conclusions. Recall that, for $\sigma_c$ nearing infinity, the two-photon spectral distribution is separable (as we assumed in our initial calculations) and each part of the two-photon state is spectrally pure. In this limit, the expression simplifies to \begin{equation}\lim\limits_{\sigma_c\to\infty}\Delta\nu_{\scriptscriptstyle SFG}\approx\frac{\ln{4}}{A}\sqrt{\left(\frac{1}{\Delta\nu_P^2}+\frac{1}{\Delta\nu_L^2}\right)}.\end{equation} This expression is identical to equation (2) of the main text, where a spectrally pure input photon was assumed.

Now we consider the limit of an infinitely narrow pump, where $\sigma_c$ approaches zero. By inspection, the only chirp-dependent term in the spectral width of equation (13) is found to be also proportional to $\sigma_c^2$. Thus, when the correlations are perfect, the spectral width is independent of the chirp parameters and can be written as \begin{equation}\lim\limits_{\sigma_c\to0}\Delta\nu_{\scriptscriptstyle SFG}=\sqrt{\Delta\nu_P^2+\Delta\nu_L^2}\end{equation} showing that compression via pulse shaping is rendered completely ineffective in the perfectly correlated limit. Note that this is the result obtained for separable states with zero chirp. Thus energy-time entanglement significantly impacts the degree of spectral compression.

The preceding calculation shows that energy-time entanglement
between the signal and idler photon can dramatically alter the
degree of bandwidth compression achieved when the idler is
traced out, i.e., ignored.  We can consider a related scenario
where the idler photon frequency is measured with outcome,
$\nu_i^0$.  If we use equation (9) as our initial state again, the
measurement of the idler frequency will leave the signal in the
pure state,
\begin{eqnarray}
|\psi\rangle=\int d\nu_s f_P(\nu_s,\nu_i^0)|1_{\nu_s}\rangle.
\end{eqnarray}
Since the signal is left in a pure state with no complications
due to entanglement, we can apply the theory from the Section I of this supplementary material. Specifically, we expect the upconverted photon to be
described by the state in equation (2).

Using the specific functional form for $f_P$ in equation (12), we
find that the single photon spectral amplitude function
$f_P(\nu_s,\nu_i^0)$ has an RMS spectral width,
$\sigma_{\mathrm{eff}}$ where,
\begin{eqnarray}
\sigma_{\mathrm{eff}}^2=\sigma^2\left(\frac{\sigma_c^2}{\sigma^2+\sigma_c^2}\right).
\end{eqnarray}
This new effective bandwidth
$\sigma_{\mathrm{eff}}^2<\sigma^2$.  A larger chirp parameter,
$A$, is thus required to reach the strong chirp limit and the
bandwidth compression in that limit (equation (5)) will be reduced.

We can again examine the extreme limits. In
the separable limit, $\sigma_c\rightarrow\infty$, measurement
of the idler photon frequency gives no information about the
signal frequency and $\sigma_{\mathrm{eff}}^2=\sigma^2$.  In
this case, we expect no modification from the bandwidth
compression predicted in Section I.  In the very
strong entanglement limit, $\sigma_c\rightarrow 0$, the
effective bandwidth of the signal
$\sigma_{\mathrm{eff}}^2\rightarrow 0$. Substituting
$A_L=-A_P=A$ and $\Delta\nu_P=0$ into equation (4), gives
$\Delta\nu_{SFG}=\Delta\nu_L$, i.e., the chirp has no impact on
the final bandwidth and the SFG bandwidth is set by that of the
laser; there is no bandwidth compression in the very strong
entanglement limit.

\section{Effect of bandwidth compression on energy-time entanglement}

We have shown that energy-time entanglement impacts the degree
of bandwidth compression.  It is interesting to turn this
question around and consider how energy-time entanglement
between the photons is modified through the nonlinear
interaction.  It is well-known that entanglement cannot
increase through local unitary interactions.  However, here we
consider only those components of the final state in which the
photon is up-converted, so entanglement could be concentrated
in a similar manner to the Procrustean protocol \cite{procrustean}.

For pure two-photon states, entanglement can be quantified
using the entropy of the reduced density matrix of either
photon. Larger entropy indicates more entanglement. The
R\'{e}nyi $\alpha$-entropy for a quantum state $\rho$ is
defined,
\begin{eqnarray}
S_\alpha(\rho) = \frac{1}{1-\alpha} \ln \mathrm{Tr} \rho^\alpha,
\end{eqnarray}
where $\alpha>1$ \cite{quantum_entropy}. In
particular, we consider the case $\alpha =2$ as it shows states
with larger entropy, and hence larger entanglement, have lower
purity, $\mathrm{Tr}\rho^2$.

Starting with the initial two-photon state from equation (9),
it can be shown that the purity can be calculated from the
reduced density matrix of the signal photon, $\rho_S$,
\begin{eqnarray}
\mathrm{Tr}_S \rho_S^2 &=& \iiiint d\nu_s d\nu_s^\prime d\nu_i d\nu_i^\prime \nonumber \\
&& f_P(\nu_s,\nu_i)f_P^*(\nu_s^\prime,\nu_i)f_P(\nu_s^\prime,\nu_i^\prime)f_P^*(\nu_s,\nu_i^\prime)\\
&=& \frac{\sqrt{\sigma_c^2(\sigma^2+\sigma_c^2)^2(2\sigma^2+\sigma_c^2)}}{(\sigma^2+\sigma_c^2)^2}.
\end{eqnarray}
Checking limits, we find that for $\sigma_c\rightarrow\infty$,
$\mathrm{Tr}_S \rho_S^2=1$ as expected for separable states,
and for $\sigma_c\rightarrow 0$ with finite $\sigma$,
$\mathrm{Tr}_S \rho_S^2=0$ as expected for perfectly entangled
continuous-variable states.

A similar calculation can be done for the idler-SFG photon
state in equation (10).  We can define the effective two-photon
spectral amplitude,
\begin{eqnarray}
\widetilde{F}(\nu_s,\nu_{SFG})=\int d\nu_i F(\nu_s,\nu_i,\nu_{SFG}).
\end{eqnarray}
Assuming the chirp rates are balanced and opposite,
$A_P=-A_L=A$ and using $\sigma_L$ to represent the laser field
RMS bandwidth and the two-photon spectral amplitude from equation (12), we calculate the purity, which is given by the complicated
expression,
\begin{widetext}
\begin{eqnarray}
\mathrm{Tr}_{SFG} \rho_{SFG}^2&=&\frac{\sigma^4\sigma_c^2(2\sigma^2+\sigma_c^2) + \sigma^2(\sigma^2+\sigma_c^2)(2\sigma^2+\sigma_c^2+4A^2\sigma^4\sigma_c^2)\sigma_L^2+4A^2\sigma^4\sigma_c^2(2\sigma^2+\sigma_c^2)\sigma_L^4}{\sqrt{\sigma^4(\sigma^2+\sigma_c^2)BC}}\\
B&=&\sigma^2\sigma_c^2(2\sigma^2+\sigma_c^2)+(\sigma^2+\sigma_c^2)(2\sigma^2+\sigma_c^2+4A^2\sigma^4\sigma_c^2)\sigma_L^2+4A^2\sigma^2\sigma_c^2(2\sigma^2+\sigma_c^2)\sigma_L^4\\
C&=&\sigma_c^2\sigma_L^2+\sigma^4\left[1+4A^2\sigma_L^2(\sigma_c^2+\sigma_L^2)\right]+\sigma^2\left[2\sigma_L^2+\sigma_c^2(1+A^2\sigma_L^4)\right].
\end{eqnarray}
The difference in the purity before and after the upconversion
process is,
\begin{eqnarray}
\mathrm{Tr}_{SFG} \rho_{SFG}^2-\mathrm{Tr}_S \rho_S^2 = \frac{\sigma^4(2\sigma^2+\sigma_c^2+4A^2\sigma^4\sigma_c^2)\sigma_L^2}{(\sigma^2+\sigma_c^2)^2\{\sigma_c^2\sigma_L^2+\sigma^4\left[1+A^2\sigma_L^2(\sigma_c^2+\sigma_L^2)\right]+\sigma^2\left[2\sigma_L^2+\sigma_c^2(1+A^2\sigma_L^4)\right]\}}
\end{eqnarray}
\end{widetext}
This quantity is nonnegative and the purity cannot decrease,
thus the entanglement cannot increase in the scenario
considered here. It is an interesting open question whether
such a statement could be generalized to higher orders of
perturbation theory. It is also interesting to see if more complex shaped pulses could enable entanglement concentration on the upconverted subensemble.

\section{ERROR ESTIMATION FOR THE SPECTRAL WIDTHS}
We estimated the uncertainties in the bandwidth of the SFG for
the single photon and weak coherent state in the following way.
We characterized the resolution of our spectrometer using a
narrow band diode laser (Toptica Bluemode) with 5 MHz of
spectral width centered at a wavelength 404 nm. Using a fit to
a Gaussian function, we measure the spectrometer resolution to
be $(0.033\pm0.002)$~nm at the diode laser wavelength. We
measured the bandwidth for several input intensities and found
that bandwidth measured has some intensity dependence at low power; this is
the dominant source of error in the bandwidth. Using the
expression $\Delta\nu=\frac{c}{\lambda^2}\Delta\lambda$ we
determine the resolution to be $\Delta\nu_R=(60\pm4)$ GHz.

We measured the spectrum of the upconverted single photons 6
times, fit each to a Gaussian function, and found the width,
$74\pm1$ GHz (see Fig. 2 in main text), where the uncertainty
is the standard deviation in the widths from the fits. To
include the spectrometer resolution we add $\pm4$ GHz in
quadrature to obtain the reported result, $\Delta\nu_M=74\pm4$
GHz. To deconvolve the spectrometer resolution from this result
for comparison with theory, we assume Gaussian spectra and use
the relation,
$\Delta\nu_{Real}=\sqrt{\Delta\nu_M^2-\Delta\nu_R^2}=43$ GHz
and Gaussian error propagation gives an uncertainty of 9 GHz.

We apply the same method for the weak coherent state. The
measured width was fitted using one spectrum only, with a value
of $67.2\pm0.4$ GHz. Including the spectrometer resolution
gives $\Delta\nu_M^{WCS}=67\pm4$ GHz, and deconvolution yields
$\Delta\nu_{Real}^{WCS}= 30\pm12$ GHz.

\section{LASER REPETITION RATE STABILIZATION}
Our experiment requires that two short pulses of light, one from the
single photon and the other from the strong laser, overlap in space
and time at the nonlinear up-conversion crystal.  Temporal jitter
between the pulses will lead to broadening of the second harmonic
spectrum since the central frequency of the light is dependent on
their relative delay.  In many scenarios, perfectly time
synchronized pulses are created by splitting a single laser pulse on
a beamsplitter.  This is common to many types of interferometry or
pump-probe techniques.  In these cases, the signal is insensitive to
changes in the repetition rate of the laser and one need only
consider small changes in the relative path lengths.  However, this
is not the case here, as the single photon (or weak coherent state)
and strong classical laser beam originate from different laser
pulses.

\begin{suppfig}
\centering
  \includegraphics[width=0.9\columnwidth]{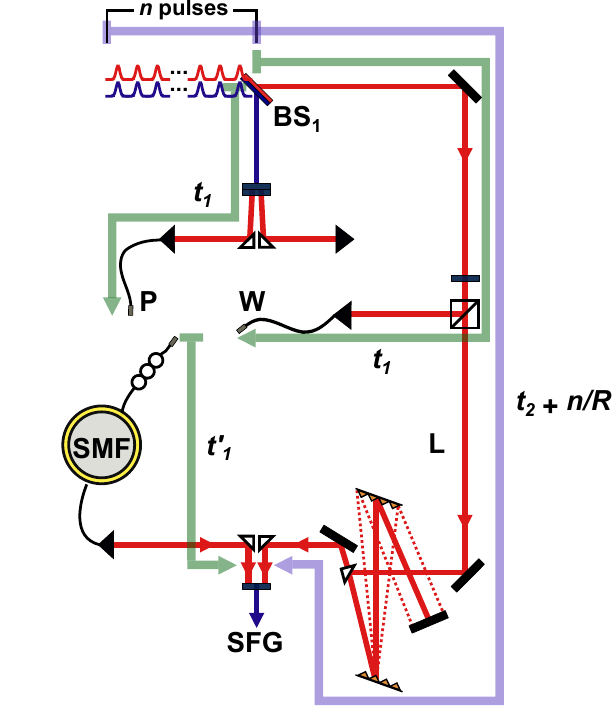}
  \caption{\textbf{Simplified version of the experimental setup.} The path length from
  the first pulse to the nonlinear crystal has an optical delay of $t_1+t_1'$.  This delay is
  equal to that experienced by the weak coherent state to the nonlinear crystal.
  The delay for the optical path from the beamsplitter to the nonlinear crystal through
  the grating-based setup is $t_2$, but since the relevant pulse originates later than that
  which creates the down-converted photon, we have to add an additional contribution
  $n/R$ where $n$ is the number of pulses later and $R$ is the repetition rate of the laser.\label{sup_setup}}
\end{suppfig}

Our experimental setup is shown in Supplementary Fig.~\ref{sup_setup}. The optical
delay for the down-conversion path is $t_1$ measured from the
beamsplitter (BS$_1$) through the down-conversion to the end of the 2 m fiber. This delay was carefully matched to that experienced by the
weak coherent state which we also label as $t_1$. The additional delay $t_1'$ corresponds the one taken by either the single photon or the weak coherent state through the 32 m of single mode fiber, up to the nonlinear crystal. The optical delay
for the strong classical pulse is $t_2$ measured from the same
beamsplitter, through the grating-based stretcher and to the
nonlinear crystal. The time delay between subsequent pulses is
$1/R$, where $R$ is the repetition rate of the laser, in our case
~80 MHz.  To this second path, we add the extra delay $n/R$ to
account for the fact that the light arriving at the nonlinear
crystal from path 2 originated $n$ pulses later than that from path
1.  Thus, the timing difference, $\Delta t$ is
\begin{eqnarray}
\label{timedifferencerelation} \Delta t = t_2 + n/R -t_1-t_1'.
\end{eqnarray}
Thus the relative time delay between either the single photon and
the strong classical pulse or the weak coherent state and the strong
classical pulse depends on both the repetition rate and the number
of pulses separating them.

\subsubsection{Experiment 1}

We measured the spectral line of the sum-frequency generated from
the weak coherent state and the strong laser pulse, as the
repetition rate was manually detuned from 80 MHz.  The results from
our experiment are shown in Supplementary Fig.~\ref{peak_vs_reprate}.  The data is
well fit by a straight line with slope $0.1188\pm0.0004$ nm/kHz.

\begin{suppfig}
\centering
   \includegraphics[width=1\columnwidth]{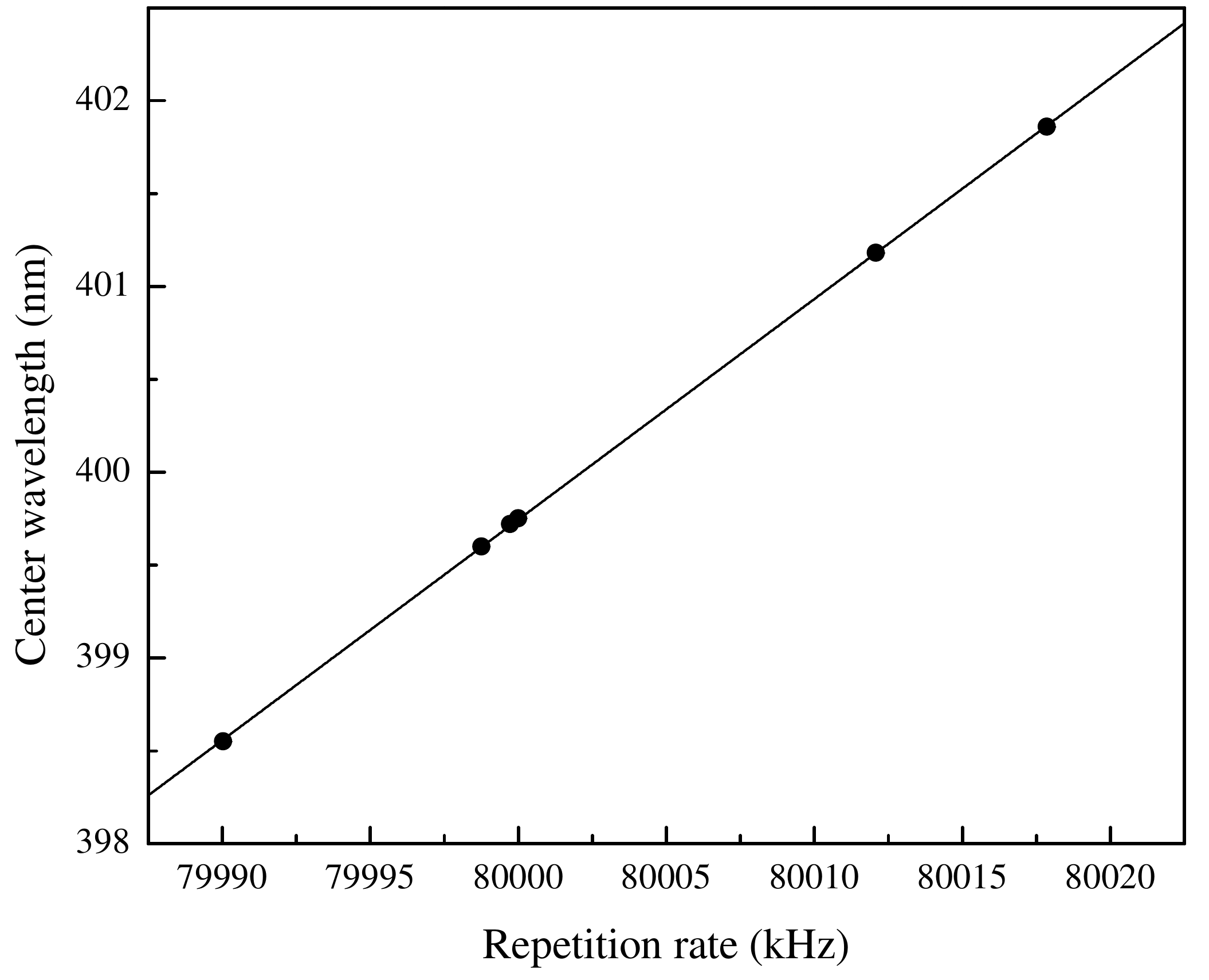}
  \caption{\textbf{Centre wavelength of the sum-frequency generation as a
 function of the laser repetition rate.} The single photon was replaced by the weak coherent state to perform this experiment. The error bars are smaller than the data points.
 \label{peak_vs_reprate}}
\end{suppfig}

To find the expected frequency change for a given variation in the
repetition rate, we start by finding the expected change in the time
delay with respect to the repetition rate
\begin{eqnarray}
\label{changeindelaywithrespecttoreprate} \frac{d(\Delta t)}{dR} =
-\frac{n}{R^2}.
\end{eqnarray}

Recall that the chirped pulses are created by applying a quadratic
frequency dependent phase, $\phi(\nu)=A(\nu-\nu_0)^2$.  The
chirp rate, the rate of change of the instantaneous frequency with
time, is $\pi/A$.  Thus we expect that a change in time of $\Delta t$
will cause a change in the frequency $d\nu = \frac{\pi}{A} d\Delta t$.
We can then convert equation~(\ref{changeindelaywithrespecttoreprate}) to
\begin{eqnarray}
\frac{d \nu}{dR} &=& -\frac{n\pi}{A R^2}\\
\frac{d \lambda}{dR} &=& \frac{\lambda^2 n\pi}{c A R^2}.
\end{eqnarray}
Substituting $\lambda = 400$ nm, $A=26.2\times10^6$ fs$^2$, $R=80$ MHz
gives
\begin{eqnarray}
\frac{d \lambda}{dR} \sim  0.01  n \frac{nm}{kHz}
\end{eqnarray}
We expect this to become significant in our experiment when the
change in wavelength approaches that of the linewidth of our single
photon, ~0.04 nm.  For $n=12$, this occurs when $\Delta R = 300$ Hz.  With
the repetition rate stabilization feature on our Spectra Physics
Tsunami (\emph{Lok-to-Clock}), this is limited to $\Delta R < 10$Hz and does not
constitute a significant source of spectral broadening.

 To estimate $n$ in our experiment, we use the slope of the center wavelength versus the relative delay, and the one of the center wavelength in function of the repetition rate. From Fig.3 in the letter, we have $\frac{d\lambda}{d t_1'}=(-0.0641\pm0.0001)$ nm/ps $=\frac{d\lambda}{dt_3}$, where $t_3=t_2+n/R.$ Additionally, we have that $\frac{d\lambda}{dR}=(0.1188\pm0.0004)$ nm/kHz from the slope of Supplementary Fig.~\ref{peak_vs_reprate}. Using $\frac{d}{dR}=\frac{dt_3}{dR}\frac{d}{dt_3}$, we find $\frac{d\lambda}{dR}=-\frac{d\lambda}{dt_3}\frac{n}{R^2}$ and hence
\begin{equation}
n=\left(\frac{d\lambda}{dt_3}\right)^{-1} \left(\frac{d\lambda}{dR}\right)R^2= 11.9\pm0.1.
\end{equation}
\noindent As expected from the lengths in our experiment, the strong laser pulse arriving at the nonlinear crystal from path 2 originates 12 pulses later than the single photon from path one.

\subsubsection{Experiment 2}

\begin{suppfig}
  \begin{center}
   \includegraphics[width=1\columnwidth]{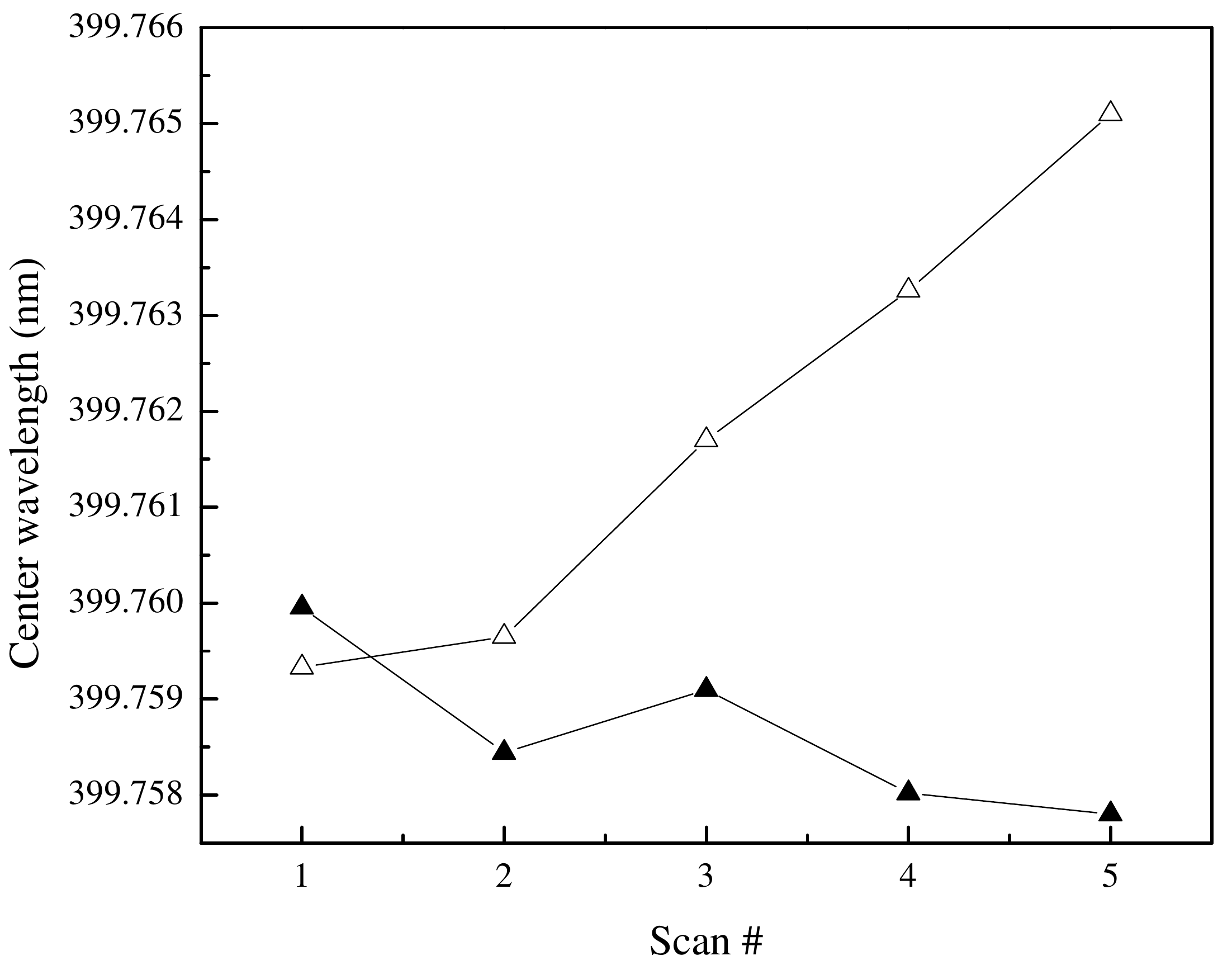}
  \end{center}
 \caption{\textbf{Lok-to-Clock feature off.} The filled triangles correspond to the case when the locking feature is on, and the empty triangles when the feature is not activated. Each scan is 10 minutes long, and repeated five
 times. The shift in the central position in the case without the Lok-to-Clock is caused by a 50 Hz detuning. The error bars are smaller than the data points.\label{laser_unlocked}}
\end{suppfig}

Throughout the experiments described in the main body of the paper
and above in the supplementary material, we employed the repetition
rate locking feature on our Titanium:sapphire laser to maintain a
constant repetition rate. We took two sets of data, one with stabilization on, the other with it off.
The results are shown in Supplementary Fig.~\ref{laser_unlocked}. For each set, we measured the peak position of the upconverted weak coherent state, with a 10 min. acquisition time and repeated five times.
The data show that the center wavelength drifts significantly without stabilization.


\end{document}